\documentclass{article}
\usepackage[utf8]{inputenc}
\usepackage{amsmath}
\usepackage{hyperref}
\usepackage{graphicx}
\usepackage{subcaption}
\usepackage{amssymb}
\usepackage{listings}
\usepackage{amsthm}
\usepackage{algorithm}
\usepackage[noend]{algpseudocode}
\usepackage{comment} 
\usepackage{amsmath,amssymb}
\usepackage{amsthm}
\usepackage{changepage}
\usepackage{cite}
\usepackage{nameref}
\usepackage{graphicx}
\usepackage{subcaption}
\usepackage{array}
\usepackage{authblk}
\usepackage{multirow}
\usepackage{xcolor}
\usepackage{color}

\usepackage{amsthm}
\usepackage{amssymb}

\theoremstyle{definition}

\theoremstyle{remark}

\usepackage[margin=0.9in]{geometry}

\usepackage{comment} 

\newcommand{\method}{CGMM}
\newcommand{\methodfull}{cancer-inspired genomics mapper model}

\title{Cancer-inspired Genomics Mapper Model for the Generation of Synthetic DNA Sequences with Desired Genomics Signatures}
\author[1*]{Teddy Lazebnik}
\author[2]{Liron Simon-Keren}
\affil[1]{Department of Cancer Biology, Cancer Institute, University College London, London, UK}
\affil[2]{School of Mechanical Engineering, Tel Aviv University, Israel}
\affil[*]{Corresponding author: t.lazebnik@ucl.ac.uk}

\date{}

\begin{document}

\maketitle

\begin{abstract}
Genome data are crucial in modern medicine, offering significant potential for diagnosis and treatment.  Thanks to technological advancements, many millions of healthy and diseased genomes have already been sequenced; however, obtaining the most suitable data for a specific study, and specifically for validation studies, remains challenging with respect to scale and access. Therefore, \textit{in silico} genomics sequence generators have been proposed as a possible solution. However, the current generators produce inferior data using mostly shallow (stochastic) connections, detected with limited computational complexity in the training data. This means they do not take the appropriate biological relations and constraints, that originally caused the observed connections, into consideration. To address this issue, we propose \methodfull{} (\method{}), that combines genetic algorithm (GA) and deep learning (DL) methods to tackle this challenge. \method{} mimics processes that generate genetic variations and mutations to transform readily available control genomes into genomes with the desired phenotypes. We demonstrate that \method{} can generate synthetic genomes of selected phenotypes such as ancestry and cancer that are indistinguishable  from real genomes of such phenotypes, based on unsupervised clustering. Our results show that \method{} outperforms four current state-of-the-art genomics generators on two different tasks, suggesting that \method{} will be suitable for a wide range of purposes in genomic medicine, especially for much-needed validation studies. 
\end{abstract}

\section{Introduction}
\label{sec:intro}
Genetic data plays a central role in the new age of medicine, providing promising abilities in both diagnostics and treatment, thanks to advances in genetic sequencing technology \cite{intro_1,intro_2,intro_3}. However, current practices in medical genetics rarely allow robust genomics analysis, due to an insufficient amount of available data \cite{intro_4}. 

Multiple computational methods have been developed to analyze genomics datasets in order to explore germline and somatic genetic variants that are associated with phenotypic traits and diseases in a population \cite{alexandrov_like_1,alexandrov_like_2,alexandrov_like_3}. While the scientific community, funding bodies, companies, and the public, recognize the importance of free and open access to genomics data for scientific research and medical progress \cite{free_data_1,free_data_2}, in practice, there is insufficient or overly complicated sharing of genomics data and associated metadata \cite{intro_problem_statement}. Several projects have attempted to address this challenge. For example, the Personal Genome Project (PGP) \cite{pgp} makes all its data available under open access. The Cancer Genome Atlas (TCGA) \cite{tcga} and the International Cancer Genome Consortium (ICGC) \cite{icgc} make different parts of their data available under two tiers, open and controlled access. Moreover, lack of data diversity (e.g. underrepresentation of rare diseases, of populations of non-European-descend and of ethnic minorities) hampers the development of many phenotype- and population-specific models required to make genomic medicine as egalitarian as possible.

To address these limitations, three main solutions have been developed. First, Data Simulators that simulate, \textit{in silico}, genotype and phenotype data \cite{in_silico_tool_1,in_silico_tool_2,in_silico_tool_3,in_silico_tool_4}. While these tools provide a solution for the lack of data problem, they do not adequately reflect the complexity of the data. Second, Data Generators generate the same type of data using more advanced statistical methods to extract genomic signatures but largely result in only slightly altered versions of the input data \cite{tool_1,tool_2,tool_3}. Both methods lack mathematical formalization limiting the level of complexity they can handle. This is mainly due to the small parameter space considered and the limited size of the feasible operations these models utilize.

Another type of genomic generative method using machine learning (ML) approaches to extract complex patterns, which are later used to generate genome sequences that satisfy these patterns \cite{ds_in_genetics_1,ds_in_genetics_2,ds_in_genetics_3}. These methods can learn extremely complex patterns and signatures in the data, however, they are often inapplicable as they require a large input dataset to work properly (which is typically not available).

To tackle these issues, we developed \method{}, a novel mathematical framework for generating synthetic genome sequences, even from scarce input data. Unlike other models, that aim to capture signatures in a single set, our approach focuses on capturing the evolutionary and mechanistic dynamics of how mutations and variants modulate a control genome into a case genome. This way, our \methodfull{} (\method{}), is able to exploit the availability of a large number of genomes as a control set and learn to map it to a small case set of genomes with the desired phenotype. The model is then able to apply the mapping it learned to convert independent control genomes into realistic case genomes. 

Specifically, \method{} is built upon a genetic algorithm that optimizes the mapping of control genomics samples into case genomics samples. The genetic algorithm uses a fitness function that is inspired by the evolutionary mechanisms that drive cancer progression, including mutation, selection, and adaptation. The fitness function incorporates multiple features of the genomic data, including sequence alignment, quality scores, and structural variations, to accurately map the synthetic reads. Afterward, \method{} incorporates a machine learning component that uses a recurrent neural network to learn and predict the optimal mapping mutation steps. The neural network is trained on a large dataset of simulated mutation patterns to learn the relationship between the genomic features and the optimal mapping parameters.

We evaluated \method{} on two datasets, demonstrating its ability to reliably multiply the desired sample count, by generating synthetic genomes associated with the case set. We compare the resulting synthetic samples to those produced by four current state-of-the-art genome generative models, demonstrating that \method{} outperforms them. 

The rest of the paper is organized as follows. In Section.~\ref{sec:related_work}, we provide an overview of the latest computational and mathematical models for \textit{in silico} DNA generation, including the relevant information inspiring the construction of \method. Section.~\ref{sec:model}, formally introduces \method. In Section.~\ref{sec:results}, we describe two \textit{in silico} experiments to evaluate the performance of \method{} compared to other genomics generator models. Finally, Section.~\ref{sec:discussion} concludes the results and proposes possible future work.

\section{Related Work}
\label{sec:related_work}
This section reviews the latest methods used to generate synthetic DNA sequences, as well as the natural evolution patterns of cancer cells and the advantages of GA in bioinformatics (which \method{} mimics and implements).

\subsection{Generation of \textit{in silico} DNA sequences}
The generation of \textit{in silico} DNA sequences based on diverse assumptions originated from previous findings of patterns in similar pathogens and samples. In particular, one defines a distribution of Single-nucleotide polymorphism (SNP) by processing available data or requesting the user to provide them, which are then randomly sampled to generate new sequences \cite{in_silico_general_rule_2,in_silico_general_rule_3}. Moreover, multiple simulators assume a linear and independent connection between the SNPs \cite{in_silico_tool_3,in_silico_general_rule_4}. For instance, \cite{in_silico_general_rule_1} proposed multi-trait, multi-locus phenotype simulations in quantitative genetics, based on a linear mix of models in a mathematical framework that takes into consideration four components: genetic variant effects, infinitesimal genetic effects, non-genetic-covariate effects, and observational noise effects as properties of \(N\) samples and \(p\) traits. This approach allows the simulator to capture realistic covariate structures. However, the model is limited by its linear mathematical nature.

Such assumptions are known to be invalid in practice since biologically, the SNPs are not randomly sampled across the genome and unnecessarily have a linear and independent relationship between them \cite{non_linear_biology_1,non_linear_biology_2,non_linear_biology_3}. For example, \cite{non_linear_biology_example} show a non-linear accumulation of structural and numeric chromosomal aberrations as well as of gene mutation during cancer progression. Therefore, several models relax these assumptions to obtain a more accurate representation of the genomics dynamics \cite{sim_sample_2,good_sim_non_linear_1,good_sim_non_linear_2}.  

\cite{good_sim_1} proposed an epistasis simulation pipeline that can generate dichotomous, categorical, and quantitative phenotypes which can simulate non-linear interaction in the distribution of SNPs while understanding observation bias. The simulation is constructed from four sequential computational processes: generating a genotype corpus, subsumption of SNPs sets, applying the epistasis model to the sampled SNPs sets, and subsampling the set of individuals. One limitation of the model is that it requires the user to declare the set of SNPs one wants to simulate specifically.

\cite{good_sim_2} presented a simulator that aims to balance the realism and computation speed for genotype data. In particular, the simulator is based on randomly sampling genetic markers as quantitative trait nucleotides (QTNs) from the whole genome or user-specified genomic area in either a normal or a geometry distribution. The proposed model allows dynamic changes in the QTNs due to the previous allocation of QTNs in different areas of the genome. This dynamic constraint allocation allows complex, non-linear dynamics but requires knowing them during the design phase rather than extracting them from the inputted data.

\cite{sim_sample_1} developed a simulator that simulates GWAS phenotype data based on user-supplied genotype data for a population and considers several biological properties such as heritability, dominance, population stratification, and epistatic interactions between SNPs. The authors divided the interaction between the SNPs into a causal sampling of the one-dimensional distribution of SNPs, and the latter in the pipeline enriched these dynamics by computing the inter-SNPs interactions. Afterward, the simulator introduces phenotype habitability in the population and population stratification to obtain case and control phenotype genomes. 

\subsection{Natural mutation process in cancer cells}
Cancer is a complex disease that arises from the accumulation of genetic and epigenetic alterations in cells that drive uncontrolled proliferation and invasion \cite{cancer}. The evolution of cancer cells from healthy ones is a multi-step process that involves the acquisition of various mutations in genes that regulate critical cellular functions, including cell cycle progression, DNA damage repair, and cell death \cite{cancer_process_1,cancer_process_2}. Once a cancer cell acquires the initial mutations, it undergoes clonal expansion and further accumulates additional mutations, leading to the emergence of subclones with distinct phenotypes and genetic profiles \cite{cancer_change}. Thus, from the cancer cells' population point of view, the amount of mutations is an ever-increasing quantity, if not treated. During the mutation process, the healthy cells that become cancer cells are bounded in the way they mutate, as unwanted mutations that are detectable by the immune system would be removed from the body \cite{cancer_bound_1,cancer_bound_2}. In a similar manner, cells that mutate at a high enough rate in a short period of time are also usually detected by the immune system and removed from the body \cite{cancer_bound_3}. In total, cancer cells increase their mutation rate during the process while being bounded in each mutation step. In our model, we utilize this evolutionary approach in the form of an accelerated-bounded mutation operator as part of a genetic algorithm scheme. 

\subsection{Genetic algorithms for bioinformatics}
A genetic algorithm (GA) is an optimization method inspired by the evolution theory \cite{ga_intro}. In particular, GA simulates the process of "evolving" through natural selection, where solutions (also referred to as "genes") that obtain a better score from the fitness function are considered more adapted. Therefore, these genes have a higher probability to pass their information to the next generation. This evolution jump between one generation to the other is performed by three stochastic processes: mutation \cite{cross_over}, crossover \cite{selection_operator}, and feasibility test \cite{feasibility_operator}. 

GAs have been used in multiple fields such as engineering \cite{ga_engineering}, medicine \cite{ga_medicine}, and economy \cite{genetic_cost_optimization}.  \cite{feasibility_operator} tackled the task of process planning optimization that is based on sequences of machines and the operations they perform. The authors used GA to obtain feasible processes for a start and later found the optimal process from the set of feasible processes. \cite{selection_operator} explored and analyzed the usage of GA with various constraints in process route sequencing and astringency. The authors reconstructed the GA, including the establishment of the coding strategy, the evaluation operator, and the fitness function, showing that the new GAs can meet the requirement of sequencing work and can meet the requirement of astringency.

In the context of genomics data, several bioinformatic challenges have been tackled using GA. For instance, \cite{ga_dna_1} show promising results studying different genetic algorithm operators for the assembly of Deoxyribonucleic acid (DNA) sequence fragments into a consensus sequence corresponding to the parent sequence, from a parent clone whose sequence is unknown. Additionally, \cite{ga_dna_2} analyzed gene expression for several cancer types, including ovarian, prostate, and lung cancer. The authors proposed an integrated gene-search algorithm for gene expression data analysis, that is based on the GA approach with correlation-based heuristics for data preprocessing. This algorithm obtained \(89\%\) accuracy in classifying 44 DNA samples into three types of cancer.

Overall, GA has demonstrated promising results in bioinformatics research \cite{ga_summary_end_1,ga_summary_end_2,ga_summary_end_3,ga_summary_end_4},and especially in genomics-related tasks \cite{ga_summary_1,ga_summary_2}. Thus, GA poses a promising computational approach to the task of generating synthetic DNA sequences.

\section{Model Definition}
\label{sec:model}
\begin{figure}[!ht]
    \centering
    \includegraphics[width=0.99\textwidth]{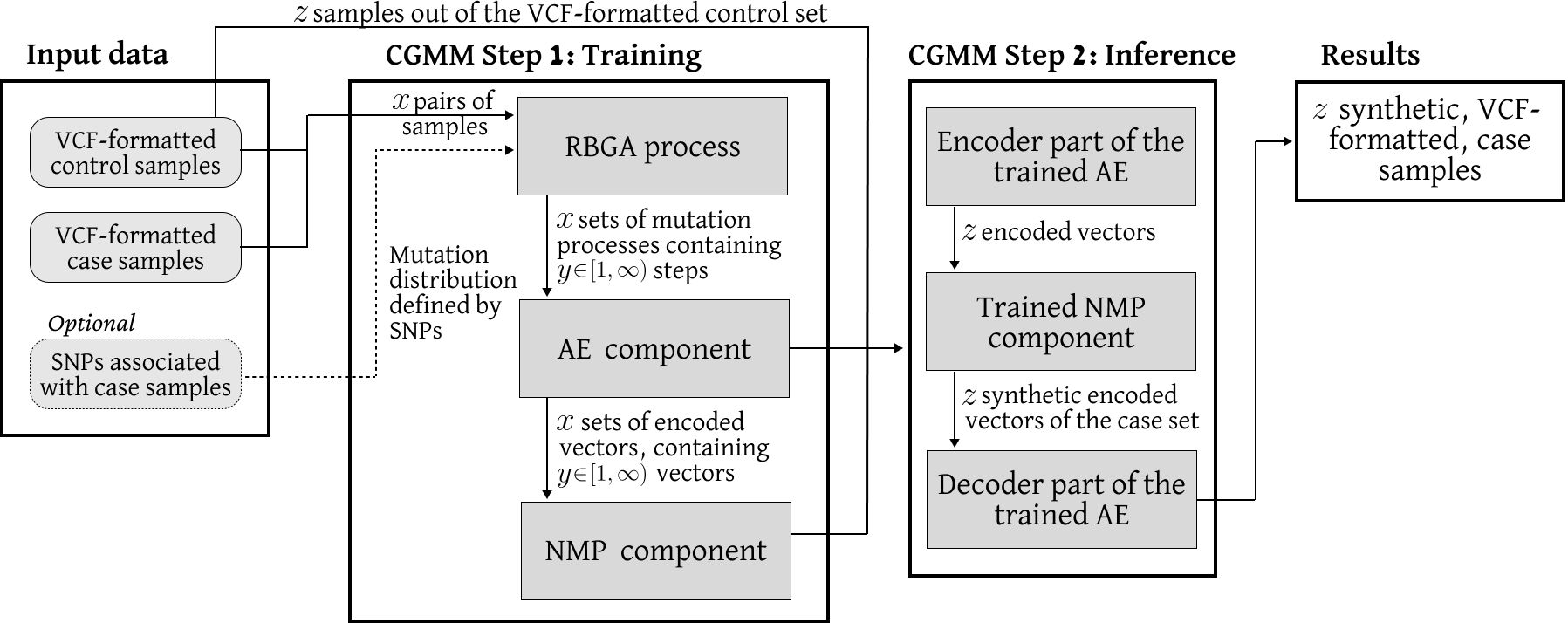}
    \caption{A schematic view of the model's training and inference phases. First, the input data is divided, leaving a specified part of the \textit{control} samples (\(z\) samples) for inference. The rest of the input data proceeds into the training phase, where: 1) The evolution processes linking each pair of \textit{control} and \textit{case} samples are estimated (RBGA process). 2) Each step in the process is encoded to a smaller, \textit{latent space} (AE component). 3) A predictive model is trained to predict the next encoded step of an evolution process (NMP component). During the inference phase, the \(z\) \textit{control} samples are used to generate encoded vectors of the first step in an evolution process (Encoder part). Then, the trained NMP component predicts a series of mutation steps, until reaching the desired genomic signatures. Finally, the last predicted step, representing an encoded, synthetic, \textit{case} sample, is decoded into VCF format (Decoder part).}
    \label{fig:training}
\end{figure}

The usage of \method{}, like many other data-driven models, is divided into two phases: training and inference. During the training phase, \method{} solves several optimization tasks to obtain the weights of inner machine learning components that construct \method. Afterward, \method{} accepts new, previously unseen,  samples as part of the inference phase and makes its prediction. A schematic view of \method{} is provided in Fig.~\ref{fig:training}, outlining the main computational steps in both the training and inference processes. 

\textbf{During the training phase}, \method{} performs four main steps. The first step is acquiring the necessary inputs of a \textit{control} set and a \textit{case} set of Variant Call Format (VCF) file format. The control and case set samples are divided in some user-defined ratio into two, where one part is for the training phase, and the second part is left for the inference phase. In addition, the user may provide SNPs associated with the case group, to be used during the training of \method. Afterward, according to some pairing approach (\(\Pi\)) chosen at the discretion of the user, \method{} generates pairs of \textit{control} and \textit{case} samples from the training dataset. Roughly divided, there are three possible pairing approaches one can take: 1) A \textit{brute-force} approach where each sample in the \textit{control} set is paired with all the samples in the \textit{case} set. This approach results in the most amount of pairs and, therefore, with the largest training dataset. However, at some point, the marginal contribution of each new pair becomes insignificant, which leads to unnecessary computations. 2) A individual-centric approach, where pairs of the \textit{control} and \textit{case} samples are obtained from the same individual. This enables the preservation of personal traits in the resulting synthetic DNA sequence containing the \textit{case} mutations. 3) A reduced \textit{brute-force} approach, where one can use any subset of the pairs group generated by the \textit{brute-force} approach. In particular, as the end goal is to learn pairing from the control to the case sets, one can take \(z \in \mathbb{N}\) random \textit{case} samples, in a uniform manner, for each \textit{control} case. While this approach introduces another level of randomness to the model, appropriate usage of if removes the disadvantages of the \textit{brute-force} approach while preserving its advantages. Therefore, this approach is the default of \method.

Once pairs of control and case samples are obtained, \method{} implements the second training step, using the \textit{Reversed bioprocess genetic algorithm} (RBGA) to generate a set of possible mutation processes from the \textit{control} to the \textit{case} samples. Formally, the RBGA is an instance of a GA algorithm with several modifications. Overall, GA follows the same computational framework with four main operators: stop condition, selection, crossover, and mutation. First, A GA would generate a random population of solution candidates. Then, the stop condition operator indicates the GA to stop once met. Otherwise, the GA would execute the other three operators in an iterative manner. The selection operator is the stage in which individual members are chosen from a population for the next iteration. This operator aims to stochastically use a fitness function to find and generate a new population of solutions that, on average, are better suited to a task in hand. The crossover operator is used to combine the information of two members to generate new members, mimicking the reproduction and biological crossover. Finally, the mutation operator introduces random changes to the members of the population. There are multiple implementations for each one of these operators that directly define the GA. For the case of RBGA, we set the fitness function to be the MASH metric \cite{mash_metric} between the member of the population and the \textit{case} sample. Following this decision, the stop condition operator is set to be \(\frac{1}{N}\sum_{i \in N} MASH(p_i, C) \leq \xi\), where \(N \in \mathbb{N}\) is the size of the GA's population which denoted by \(P\), \(p_i \in P\) is each member in the current GA's population, \(C\) is the \textit{case} sample, and \(\xi \in \mathbb{R}^+\) is a threshold value set by the user. The selection operator (also known as the \textit{next-generation} operator) is implemented to be the "\textit{tournament with royalty}" \cite{selection_operator}. We choose this selection operator as multiple studies show it performs better than other operators and is able to converse in complex and multi-dimensional optimization tasks such as ours \cite{next_gen_operator_review}. For the crossover operator, we choose at random (in a uniformly distributed manner) pairs of members from the GA's population and a pivot point, generating two new members for the GA's population. These members replace the original two by crossing the two parts of each from the original members, as defined by the random pivot point \cite{cross_over}. Lastly, we propose a novel mutation process inspired by the way cancer mutates. Similar to how cancer cells (representing \textit{case} samples) evolve from healthy cells (representing \textit{control} samples), \method{} removes, adds, or edits SNPs in the \textit{control} DNA sequence, while maintaining similarity to its origin. To do so, we adopt the decreasing learning rate approach from the adaptive gradient descent algorithm \cite{agd}, while also bounding the total distance the changes can cause in a single mutation step. Precisely, let \(\mu \in \mathbb{R}^+\) and \(\sigma  \in \mathbb{R}^+\) be the mean and standard deviation of a normal distribution stochastic variable \(\zeta \sim N(\mu, \sigma^2)\). In addition, let us define a discount factor \(\lambda \in (0, 1)\). Now, the accelerated-bounded mutation operator operates as follows. For the \(i_{th}\) generation of the GA, we choose a random number of mutations \(\zeta_i\) that distributed \(N(\mu \cdot \lambda^i, \sigma^2 \cdot \lambda^i)\). Then, we apply \(\zeta_i\) changes to the member reflected by random SNPs. If the MASH metric distance between the member after the mutation and the sample before the mutation is larger than a pre-defined threshold \(\beta \in \mathbb{R}^+\), the mutated member is considered infeasible, and the process repeats. Moreover, at this stage, if the user provided SNPs associated with the case group, the distribution of these SNPs in the case group is calculated and used as the distribution function for the mutation operator (replacing the uniform distribution function).

After a predefined number of pairs (or all of them, if not set otherwise) produce the mutation processes, \method{} initiates the third step of the training phase, in which a model of AutoEncoder (AE) with a self-attention layer is trained. Formally, the AE accepts a mutation process in the form of a list of numbers, which indicate the index in the genome array and one of the 12 possible SNP  mutations. The AE is generated using the AutoKeras automatic deep learning library, to find the near-optimal AE architecture \cite{autokeras} for the specific inputs inserted. To do so, the AutoKeras framework aims to minimize the mean square error between the input and the reconstructed samples, using Bayesian optimization on the size of the encoding layer \cite{autoencoder_tunning}. Once the AE model is obtained, each step in the mutation processes generated during the RBGA is converted into an encoded vector. This results in a set of vectors with the same dimension (i.e., the AE's encoding layer's dimension), describing a possible mutation process that links a specific pair of \textit{control} and \textit{case} samples. 

In the fourth and final step of the training phase, \method{} uses a recurrent neural network (RNN) architecture, with a long-short term memory layer (LSTM), to predict the next step in a possible mutation process between pairs of \textit{control} and \textit{case} samples. This RNN component of \method{} is denoted as the Next Mutation Predictor (NMP). The NMP works as a time-series predictor, similar to the way Large Language Models learn to predict the next word in a sentence \cite{word_embedding}. In particular, it is generated using the model architecture proposed by \cite{next_word_model}, as it showed promising accuracy on the next-word prediction task while keeping the model's size relatively small, and therefore, less computationally expensive. Like the work of \cite{next_word_model}, the NMP receives a special token indicating the mutation process is over every time it reaches the end of an encoded representation vector. This indicates to the NMP that the specific mutation process has ended and prevents it from attempting to predict the next step in the process indefinitely. 

Finally, the trained NMP and the AE models are saved for later usage in the inference phase.

\textbf{During the inference phase}, \method{} performs three main steps, this time on the input of \(\alpha \in \mathbb{N} \)\textit{control} samples that were set aside for inference. In the first step, \method{} implements the Encoder part of the AE generated during training, to encode the samples into encoded vectors. In the second step, the encoded vectors are passed to the NMP model, where their respective mutation process into \textit{case} samples are predicted. Finally, the resulting predictions are decoded in the Decoder part of the AE, converting them back into VCF format sequences. This results in a set of \(\alpha\) synthetic \textit{case} samples.   

\section{Experimental design}
We tested \method{} on three tasks, to demonstrate its capability to reliably increase the readily available DNA sequences of the \textit{case} set. To do so, we collected genomics sequences from public repositories and divided them into the desired \textit{case} set, which we needed to enrich with synthetic data, and the control set of DNA sequences, that we used for this task. Specifically, for the first two experiments (Exps. 1 and 2), we used 300 samples from the Personal Genomics Project (PGP), from three countries: the United Kingdom (120 samples), Canada (110 samples), and the United States (70 samples). The samples were manually divided into 200 European-dominant DNA sequences, acting as the control set, and 100 African-dominant DNA sequences, acting as the \textit{case} set. In order to make sure the tagged ancestry is valid, we verified that the tags of all the samples correspond both to a PCA-based ancestry classification and to the ancestry self-reported by the examined individuals \cite{ansectry}. For the final experiment (Exp.3), we used 90 control samples from PGP corresponding to healthy individuals from the United Kingdom, and 30 \textit{case} samples were obtained from Genomic Data Commons (GDC), corresponding to individuals with skin melanoma. 

To demonstrate the balance between the conversion rate of control samples into \textit{case} samples, and the absolute number of successfully converted control samples, we used a 50\%-50\% training and testing ratio for Exp. 2, compared to a 70\%-30\% ratio for Exps. 1 and 3. The resulting sample counts for each phase are presented in Table.~\ref{tab:trai_test_ratio}.

\begin{table}[ht]
    \centering
    \begin{tabular}{|c|cc|cc|cc|}
        \hline
        & \textbf{Exp.1} && \textbf{Exp.2} && \textbf{Exp.3} \\
        & \textit{Contro}l & \textit{Case} & \textit{Control} & \textit{Case} & \textit{Control} & \textit{Case} \\ \hline
        Train & 140& 70 & 100 & 50 & 60 & 20 \\
        Test & 60 & 30 & 100 & 50 & 30 & 10\\
        \hline
    \end{tabular}
    \caption{The sample count of \textit{control} and \textit{case} samples used during the training and inference phases. For inference, only the \textit{control} samples are needed for \method's operation, whilst the listed \textit{case} samples are used to evaluate the reliability of the results.}
    \label{tab:trai_test_ratio}
\end{table}

Once the dataset was divided, the model was trained using the hyperparameters listed in Table.~\ref{table:parameters}, provided in the Appendix. 

To demonstrate that \method{} is capable of learning the correct relation and genomics signatures associated with each \textit{case} set, we repeated each experiment twice: once without any knowledge integration, then with outside knowledge in the form of ancestry-related SNPs from ForenSeq™ DNA \cite{snps_for_exps} (Exp.1-2), or in the form of skin-melanoma-related SNPs from \cite{alexandrov}.

To evaluate the performance of \method{}, we used hierarchical clustering following the CLINK algorithm \cite{h_clustering}, and computed the distances between samples using the MASH metric \cite{mash_metric}. While there are many other metrics, such as the Jensen-Shannon divergence (JSD) \cite{other_metric_1} and the Kullback-Leibler (KL) divergence metrics \cite{other_metric_2}, we decided to use MASH due to its popularity and well-established validation in bioinformatics. We then report the conversion rate, meaning the percentage of \textit{control} samples that were classified as \textit{case} samples out of the total \textit{control} samples used during inference. The results of \method{} for each experiment are compared to those obtained by four other models, considered as state-of-the-art genomics generators \cite{good_sim_2,similar_data_driven_1,in_silico_general_rule_1,sim_sample_1}.

\section{Results}
\label{sec:results}
Table.~\ref{table:models_comparision} presents the conversion rate of \textit{control} samples into reliable \textit{case} samples, as achieved by the five examined models. As expected, the results indicate that the more training samples obtained, the easier it becomes for the models to detect the genomic signatures associated with each \textit{case} set. Therefore, Exp.1 leads to the highest conversion rates for all models (neglecting cases where SNPs are included). However, once SNPs' knowledge is integrated, the effect of the training sample count reduces significantly. Additionally, in all experiments \method{} is the least affected by the addition of SNPs knowledge, gaining only a \(15.6\% \pm  10.2\%\) increase in conversion rates for Exp.1-3, compared to a \(36\% \pm  2.8\%\) increase observed in the rest of the models. 

\begin{table}[!ht]
\centering
\begin{tabular}{lccccc}
\hline
\textbf{Test configuration} & \textbf{\method{}} & \textbf{G2P} \cite{good_sim_2} & \textbf{Zhou et al.} \cite{similar_data_driven_1} & \textbf{PhenotypeSimulator} \cite{in_silico_general_rule_1} & \textbf{GEPSi} \cite{sim_sample_1} \\ \hline
\(E_1\)           & 78.3\% & 38.3\% & 31.6\% & 40.0\% & 56.6\%  \\ 
\(E_1\) with SNPs  & 85.0\% & 71.6\% & 68.3\% & 83.3\% & 81.6\% \\ \hline
\(E_2\)            & 73.3\% & 31.6\% & 26.6\% & 38.3\% & 41.6\%  \\ 
\(E_2\) with SNPs & 86.6\% & 68.3\% & 73.3\% & 81.6\% & 78.3\% \\ \hline
\(E_3\)         & 46.6\% & 23.3\% & 16.6\% & 36.6\% & 30.0\%  \\ 
\(E_3\) with SNPs & 73.3\% & 53.3\% & 46.6\% & 66.6\% &  70.0\% \\ \hline
\end{tabular}
\caption{Comparison of the conversion rate of \textit{control} samples into \textit{case} samples, as achieved by \method{} and four state-of-the-art models.}
\label{table:models_comparision}
\end{table}

\begin{figure}[!ht]
    \centering
    \includegraphics[width=.8\textwidth]{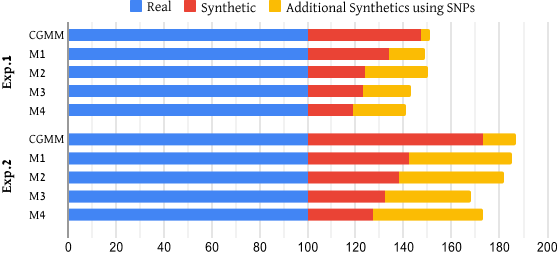}
    \caption{The resulting sample count of the \textit{case} set DNA sequences, comprised of the original, real samples (blue), the synthetic samples produced without any SNP knowledge (red), and the additional synthetic sample count that was achieved with knowledge integration (yellow). The results of \method{} are compared to those achieved by GEPSi (M1), PhenotypeSimulator (M2), G2P (M3), and Zhou et al. (M4).}
    \label{fig:absolute_increase}
\end{figure}

Fig.~\ref{fig:absolute_increase} demonstrates the increase in the sample count of the resulting \textit{case} dataset of DNA sequences in Exp.1-2. Although both experiments started with the same baseline count of samples (blue), and Exp.2 led to smaller conversion rates for all models (see Table.~\ref{table:models_comparision}), models achieved a more optimal result in Exp.2 than in Exp.1. In other words, they all resulted in a larger sum of reliable synthetic DNA sequences (either with SNPs knowledge (yellow) or without (red)). 

\section{Discussion and conclusion}
\label{sec:discussion}
In this study, we proposed a novel \methodfull{} (\method{}) that combines a genetic algorithm with a unique cancer-like mutation process, a deep learning AutoEncoder, and a next-word prediction model, to capture the non-linear dynamics linked to the mutation process between two genomics samples. Using our \method{} model, we were able to generate synthetic genomes while maintaining the personal traits of the input samples. This was done by learning the possible mutation evolution process between the pairs of samples and bounding each evolutionary step with logical assumptions that mimic the evolution of cancer cells. Thus, \method{} is able to receive an input genome and mutate it until some desired property is achieved. Therefore, our computational approach differs from current synthetic genomics models, which try to capture the unique statistical properties on the genomics level, given a dataset of genomes with a desired property (such as ancestry or pathogen). Opposed to them, we aim to capture the statistical mutation dynamics between a dataset of \textit{control} samples, that are more readily available, and a dataset of \textit{case} samples, that possess the desired property we want. This way, \method{} is able to take advantage of genome samples that are found in abundance in order to reliably enrich a different, smaller dataset of genome samples, with synthetic samples.
 
In order to evaluate \method{}'s performance compared to other synthetic genome generation models \cite{good_sim_2,similar_data_driven_1,in_silico_general_rule_1,sim_sample_1}, we conducted three experiments that aim to generate genomes with a desired property (Exp. 1-3). For that, we used two datasets, one of European and African-dominant ancestries, and the other of healthy individuals and individuals with skin melanoma. In all experiments, as shown in Table. \ref{table:models_comparision}, \method{} outperformed the other models, generating a higher conversion rate of \textit{control} samples into reliable, synthetic \textit{case} samples. 

We repeated each experiment, integrating outside knowledge in the form of SNPs. The results show that the integration of knowledge further improves the performance of \method{}, however, at a much smaller rate than that of all other models (as seen in Table.~\ref{table:models_comparision} and Fig.~\ref{fig:absolute_increase}). This leads to comparable results of \method{} and the rest of the state-of-art models if knowledge is integrated. Thus we argue that: 1) The results demonstrate that \method{}, opposed to the other models, is capable of learning the correct relation between the \textit{control} and \textit{case} set. Hence, the integration of knowledge has a relatively low effect on it. 2) \method{} seems to be able to capture SNP signatures partially. This is a promising outcome as with a proper adoption, \method{} can be used for SNP signature detection as well. 3) Without knowledge of SNP signatures, \method{} may achieve better results than the other models, even on small datasets. This is because the mutation path approach of \method{} can extend the training set to a much larger cohort, by pairing the \textit{case} samples with multiple \textit{control} samples. This, in turn, increases the number of mutation processes available for the training of \method{}, easing the deduction of synthetic \textit{case} samples.  

Additionally, Fig. \ref{fig:absolute_increase} shows that one has to balance the need for a large training set (which aids in achieving a higher conversion rate), and the need to preserve samples for the generation of synthetic ones. Thus, we show that although a higher conversion rate is achieved in Exp.1, a larger, absolute increase in sample count is achieved in Exp.2 (i.e., \method{} generated in Exp.2 31 more reliable synthetic samples than in Exp.1 although it was trained on a smaller dataset of samples).

This study has several limitations that should be addressed in future work to improve further the performance and usability of \method{}. First, the AutoEncoder component in \method{} is designed to optimize an encoding task, as obtained by using the auto deep learning approach, but does not utilize any domain knowledge that can be helpful \cite{kiae}. Second, the current method is not explainable, which may limit its usage in the clinical domain \cite{pick_tau_1,post_prunning}. Third, due to the lack of freely available genomics data, the experiments were conducted on relatively small datasets. While these experiments represent important usage configurations, a test on a statistically larger dataset can allow obtaining a more robust evaluation of \method{}.  

\section*{Declarations}
\subsection*{Funding}
This research received no specific grant from funding agencies in the public, commercial, or not-for-profit sectors.

\subsection*{Conflicts of interest/Competing interests}
The authors have no financial or proprietary interests in any material discussed in this article.

\subsection*{Data availability}
The data used is publicly available, and the sources are cited in the text.

\subsection*{Author Contributions}
Teddy Lazebnik: Conceptualization, data curation, formal analysis, investigation, methodology, software, visualization, supervision, writing - original draft, and writing - review \& editing. \\ 
Liron Simon-Keren: Formal analysis, visualization, and writing - original draft. \\

\subsection*{Acknowledgements}
The authors wish to thank Olga Chervova, Vitaly Voloshin, Ismail Moghul, and Stephan Beck for their help in this project. 

\bibliographystyle{unsrt}
\bibliography{bilbo}

\section*{Appendix}
The proposed model's hyperparameters descriptions and values have been used as part of the presented experiments. These values are chosen throughout the trial and error process during the development of \method{} for a subset of the samples used in the presented Exp 1.

\begin{table}[!ht]
\begin{tabular}{|p{0.7\textwidth}|p{0.2\textwidth}|}
 \hline
\textbf{Parameter definition}  & \textbf{Value} \\ \hline
Pairing approach [\(1\)]  & Subset approach with 10 pairs for each \textit{case} samples \\ \hline
Number of generations in RBGA [\(1\)]  & \(100\) \\ \hline
Initial mutation rate in the genetic algorithm [\(1\)]  & \(2.5 \cdot 10^{-4}\) \\ \hline
Mutation rate decrease factor in the genetic algorithm [\(1\)] & \(1 \cdot 10^{-5}\) \\ \hline
Royalty rate in the genetic algorithm [\(1\)]  & \(0.05\)  \\  \hline
The threshold value, \(\zeta\), for the stop condition  [\(1\)] & \(1.5 \cdot 10^3\)  \\  \hline
Bounding factor, \(\beta\), for the accelerated-bounded mutation operator [\(1\)] & \(4.5 \cdot 10^3 \)  \\  \hline
AutoEncoder's encoding layer dimension [\(1\)] & 8192 \\  \hline
\end{tabular}
\caption{The description of the model’s hyperparameters and their values.}
\label{table:parameters}
\end{table}

\end{document}